\documentstyle[pra,aps,multicol]{revtex}
\renewcommand{\narrowtext}{\begin{multicols}{2} \global\columnwidth20.5pc}
\renewcommand{\widetext}{\end{multicols} \global\columnwidth42.5pc}
\begin{document}
\draft
\title{{\bf Multipartite pure-state entanglement and the generalized GHZ states}}
\author{Shengjun Wu, Yongde Zhang}
\address{{\it Department of Modern Physics,}
University of Science and Technology of China, Hefei, Anhui 230027,
P.R.China}
\date{\today}
\maketitle

\begin{abstract}
We show that not all 4-party pure states are GHZ reducible (i.e., can be
generated reversibly from a combination of 2-, 3- and 4-party maximally
entangled states by local quantum operations and classical communication
asymptotically) through an example, we also present some properties of the
relative entropy of entanglement for those 3-party pure states that are GHZ
reducible, and then we relate these properties to the additivity of the
relative entropy of entanglement.
\end{abstract}


\vskip 1.0cm

\narrowtext

\section{INTRODUCTION}

Ever since it was first noted by Einstein-Podolsky-Rosen (EPR) \cite{epr}
and Schr\"odinger \cite{Sc}, entanglement has played an important role in
quantum information theory. Quantum entanglement provides strong tests of
quantum nonlocality \cite{Be,bc}, and it is also a useful resource for
various kinds of quantum information processing, including teleportation 
\cite{bbcjpw,bpz}, cryptographic key distribution \cite{des}, quantum error
correction \cite{Sh} and quantum computation \cite{De}.

Now, one of the key open questions in quantum information theory is how many
fundamentally different types of quantum entanglement there are. It was
known \cite{bbps,pr} that asymptotically there is only one kind of
entanglement for bipartite pure states, any pure entangled state of two
parties (Alice and Bob) may be reversibly transformed into EPR states by
local quantum operations and classical communication (LOCC) asymptotically.

For multipartite pure states, it is more difficult to understand the types
of entanglement. It was not known whether the EPR states are the only type
of entanglement until the recent work of
Bennett-Popescu-Rohrlich-Smolin-Thapliyal (BPRST) \cite{bpt}, which shows
that the 4-party GHZ state can not be reversibly transformed into EPR states
by LOCC asymptotically. Furthermore Linden-Popescu-Schumacher-Westmoreland
(LPSW) \cite{lpsw} had shown that the n-party GHZ state cannot be reversibly
transformed into any combination of $k$-party entangled pure states for all $%
k<n$. This means that the generalized $n$-party GHZ state 
\begin{equation}
\left| GHZ_n\right\rangle _{ABC\cdots }\equiv \frac 1{\sqrt{2}}(\left|
0^{\otimes n}\right\rangle +\left| 1^{\otimes n}\right\rangle )
\end{equation}
represents a different type of entanglement with respect to the $k$-party
GHZ state (for all $k\neq n$) \footnote{%
A particular $n$-party GHZ state is chosen to represent all the $n$-party
GHZ states since they are related by local unitary transformations.}.

A natural question arises, are the generalized GHZ states the only types of
entanglement? Thapliyal \cite{Th} had shown that any multi-separable pure
state is Schmidt decomposable, thus a $m$-party separable pure state ( a
state contains no entanglement of $k$-party for all $k<m$) can be reversibly
transformed into the $m$-party GHZ state, this result supports (but does not
prove) the hypothesis that the generalized GHZ states are the only types of
entanglement with the $k$-party GHZ state representing ''essential'' $k$%
-party entanglement.

In this note, we show that the generalized GHZ states are not the only types
of entanglement through an example of 4-party pure state. We also present
some properties of the relative entropy of entanglement for those 3-party
pure states that can be generated reversibly from 2- and 3-party GHZ states,
and then we use these properties to analyze the additivity of the relative
entropy of entanglement.

Before going to the results, we state some terminology more clearly. Two $m$%
-party pure states $\left| \psi \right\rangle $ and $\left| \varphi
\right\rangle $ are LOCCa equivalent if and only if\cite{bpt} 
\begin{eqnarray}
\forall _{\delta >0,\epsilon >0} \exists _{n_1,n_2,n_3,n_4,L,L^{^{\prime }}} 
\text{ so that }  \nonumber \\
|(n_1/n_2)-1| &<&\delta \text{ , }  \nonumber \\
|(n_3/n_4)-1|<\delta \text{ , and}  \label{def1} \\
F\left( L(\left| \psi \right\rangle ^{\otimes n_1}),\left| \varphi
\right\rangle ^{\otimes n_2}\right) &\geq &1-\epsilon  \nonumber \\
F\left( L^{^{\prime }}(\left| \psi \right\rangle ^{\otimes n_3}),\left|
\varphi \right\rangle ^{\otimes n_4}\right) &\geq &1-\epsilon  \nonumber
\end{eqnarray}
where $L$ and $L^{^{\prime }}$ are local quantum operations assisted by
classical communication, and 
\begin{equation}
F\left( \left| \Phi \right\rangle ,\left| \Psi \right\rangle \right) \equiv
\left| \left\langle \Phi \right. \left| \Psi \right\rangle \right| ^2
\end{equation}
is the fidelity of $\left| \Psi \right\rangle $ relative to $\left| \Phi
\right\rangle $. Condition (\ref{def1}) means that, in the limit of large $n$%
, $n$ copies of $\left| \psi \right\rangle $ can be transformed into almost
the same number of copies of $\left| \varphi \right\rangle $ by LOCC with
high fidelity, and vice versa. The LOCCa equivalence of the two $m$-party
pure states $\left| \psi \right\rangle $ and $\left| \varphi \right\rangle $
is denoted as 
\begin{equation}
\left| \psi \right\rangle \stackrel{LOCCa}{\rightleftharpoons }\left|
\varphi \right\rangle
\end{equation}

We say that a $m$-party pure state $\left| \psi \right\rangle $ is GHZ
reducible, if and only if the state $\left| \psi \right\rangle $ is LOCCa
equivalent to a combination of $2$-, $3$-, $\cdots $, $m$-party generalized
GHZ states. For example, since any bipartite pure state $\left| \psi
\right\rangle _{AB}$ is GHZ reducible \cite{bpt,Th}, we can write 
\begin{equation}
\left| \psi \right\rangle _{AB}\stackrel{LOCCa}{\rightleftharpoons }\left|
EPR\right\rangle _{AB}^{\otimes E(\left| \psi \right\rangle _{AB})}
\end{equation}
where $E(\left| \psi \right\rangle _{AB})$ is the unique measure of
entanglement for bipartite pure states, it is equal to the entropy of the
reduced density matrix of either Alice or Bob, as well as the entanglement
of formation \cite{hw}, entanglement of distillation \cite{bbps,bdsw} and
the relative entropy of entanglement \cite{vprk}. A $3$-party GHZ reducible
pure state $\left| \psi \right\rangle _{ABC}$ can be written as 
\begin{eqnarray}
\left| \psi \right\rangle _{ABC}\stackrel{LOCCa}{\rightleftharpoons }\left|
EPR\right\rangle _{AB}^{\otimes E_2(AB)}\otimes \left| EPR\right\rangle
_{AC}^{\otimes E_2(AC)}  \nonumber \\
\otimes \left| EPR\right\rangle _{BC}^{\otimes E_2(BC)}\otimes \left|
GHZ\right\rangle _{ABC}^{\otimes E_3(ABC)}
\end{eqnarray}
i.e., in the limit of large $n$, with high fidelity, $n$ copies of the state 
$\left| \psi \right\rangle _{ABC}$ can be transformed reversibly by LOCC
into $n\cdot E_2(AB)$ copies of the state $\left| EPR\right\rangle _{AB}$
held by Alice and Bob, $n\cdot E_2(AC)$ copies of $\left| EPR\right\rangle
_{AC}$ held by Alice and Claire, $n\cdot E_2(BC)$ copies of $\left|
EPR\right\rangle _{BC}$ held by Bob and Claire, and $n\cdot E_3(ABC)$ copies
of $\left| GHZ\right\rangle _{ABC}$ held by Alice, Bob and Claire. The GHZ
reducible multipartite pure states can be written in similar forms.

Let \c C denotes a set of pure states, if each of the $m$-party pure states
is LOCCa equivalent to a certain combination of the states in \c C, then we
say that \c C is a reversible entanglement generating set (REGS) \cite{bpt}
for $m$-party pure states. A minimal reversible entanglement generating set
(MREGS) for $m$-party pure states is a REGS of minimal cardinality. It is
obvious that the set \c C$_2$=\{$\left| EPR\right\rangle _{AB}$\} is a MREGS
for bipartite pure states. The question that whether the set of $2$-, $3$-, $%
\cdots $, $m$-party generalized GHZ states is a MREGS for $m$-party pure
states is, in fact, equivalent to the question that whether all $m$-party
pure states are GHZ reducible.

We now state BPRST's lemma about the LOCC equivalence.

{\it BPRST's lemma}: If two $m$-party quantum states $\left| \Psi
\right\rangle $ and $\left| \Phi \right\rangle $ are LOCCa equivalent, then
they must be isentropic, i.e., 
\begin{equation}
S_X\left( \left| \Psi \right\rangle \right) =S_X\left( \left| \Phi
\right\rangle \right)
\end{equation}
where $S_X\left( \left| \Psi \right\rangle \right) =-tr\left\{ \rho _X\left(
\left| \Psi \right\rangle \right) \log _2\rho _X\left( \left| \Psi
\right\rangle \right) \right\} $ with $\rho _X\left( \left| \Psi
\right\rangle \right) \equiv tr_{\overline{X}}\left( \left| \Psi
\right\rangle \left\langle \Psi \right| \right) $, and $X$ denotes a
nontrivial subset of the parties (say Alice, Bob, Claire, Daniel, et al.), $%
\overline{X}$ denotes the set of the remaining parties.

This lemma is a consequence of the fact that average partial entropy $S_X$
cannot increase under LOCC, details of proof can be found in ref. \cite
{bdsw,bpt}.

\section{The set of generalized GHZ states is not a MREGS}

Now we show that the generalized GHZ states are not the only types of
entanglement by proving that the set of $2$-, $3$-, $4$-party GHZ states is
not a MREGS for $4$-party pure states, or in another word, not all $4$-party
pure states are GHZ reducible.

{\sl Proposition 1:} The set of $2$-, $3$-, $4$-party GHZ states is not a
MREGS for $4$-party pure states.

Before the proof, let us first give a property of all the GHZ reducible $4 $%
-party pure states. Suppose the $4$-party pure state $\left| \Psi
\right\rangle _{ABCD}$ is GHZ reducible, i.e.,

\widetext
\begin{eqnarray}
\left| \Psi \right\rangle _{ABCD}&\stackrel{LOCCa}{\rightleftharpoons }%
&\left| EPR\right\rangle _{AB}^{\otimes E_2(AB)}\otimes \left|
EPR\right\rangle _{AC}^{\otimes E_2(AC)}\otimes \left| EPR\right\rangle
_{AD}^{\otimes E_2(AD)} \otimes \left| EPR\right\rangle _{BC}^{\otimes
E_2(BC)}  \nonumber \\
&&\otimes \left| EPR\right\rangle _{BD}^{\otimes E_2(BD)}\otimes \left|
EPR\right\rangle _{CD}^{\otimes E_2(CD)} \otimes \left| GHZ\right\rangle
_{ABC}^{\otimes E_3(ABC)} \otimes \left| GHZ\right\rangle _{ABD}^{\otimes
E_3(ABD)}  \label{x1} \\
&&\otimes \left| GHZ\right\rangle _{ACD}^{\otimes E_3(ACD)}\otimes \left|
GHZ\right\rangle _{BCD}^{\otimes E_3(BCD)}\otimes \left| GHZ_4\right\rangle
_{ABCD}^{\otimes E_4(ABCD)}  \nonumber
\end{eqnarray}
From BPRST's lemma and the additivity of the von Neumann entropy, we have 
\begin{equation}
\left. 
\begin{tabular}{l}
$S\left( \rho _A\right) =E_2\left( AB\right) +E_2\left( AC\right) +E_2\left(
AD\right) +E_3\left( ABC\right) +E_3\left( ABD\right) +E_3\left( ACD\right)
+E_4\left( ABCD\right) $ \\ 
$S\left( \rho _B\right) =E_2\left( AB\right) +E_2\left( BC\right) +E_2\left(
BD\right) +E_3\left( ABC\right) +E_3\left( ABD\right) +E_3\left( BCD\right)
+E_4\left( ABCD\right) $ \\ 
$S\left( \rho _C\right) =E_2\left( AC\right) +E_2\left( BC\right) +E_2\left(
CD\right) +E_3\left( ABC\right) +E_3\left( ACD\right) +E_3\left( BCD\right)
+E_4\left( ABCD\right) $ \\ 
$S\left( \rho _D\right) =E_2\left( AD\right) +E_2\left( BD\right) +E_2\left(
CD\right) +E_3\left( ABD\right) +E_3\left( ACD\right) +E_3\left( BCD\right)
+E_4\left( ABCD\right) $%
\end{tabular}
\right.  \label{a}
\end{equation}
and 
\begin{equation}
\left. 
\begin{array}{lll}
S\left( \rho _{AB}\right) & = & E_2\left( AC\right) +E_2\left( AD\right)
+E_2\left( BC\right) +E_2\left( BD\right) +E_3\left( ABC\right) \\ 
&  & +E_3\left( ABD\right) +E_3\left( ACD\right) +E_3\left( BCD\right)
+E_4\left( ABCD\right) \\ 
S\left( \rho _{AC}\right) & = & E_2\left( AB\right) +E_2\left( AD\right)
+E_2\left( BC\right) +E_2\left( CD\right) +E_3\left( ABC\right) \\ 
&  & +E_3\left( ABD\right) +E_3\left( ACD\right) +E_3\left( BCD\right)
+E_4\left( ABCD\right) \\ 
S\left( \rho _{AD}\right) & = & E_2\left( AB\right) +E_2\left( AC\right)
+E_2\left( BD\right) +E_2\left( CD\right) +E_3\left( ABC\right) \\ 
&  & +E_3\left( ABD\right) +E_3\left( ACD\right) +E_3\left( BCD\right)
+E_4\left( ABCD\right)
\end{array}
\right.  \label{b}
\end{equation}
\narrowtext
From Eqs. (\ref{a}) it follows that 
\begin{equation}
\sum_{i\in \left\{ A,B,C,D\right\} }S\left( \rho _i\right) =2\cdot
E_{2t}+3\cdot E_{3t}+4\cdot E_4  \label{c}
\end{equation}
with $E_{2t}$ ($E_{3t}$, $E_4$) representing the "total" $2$- ($3$-, $4$-)
party entanglement, which is defined by 
\begin{equation}
\left. 
\begin{array}{lll}
E_{2t} & = & E_2\left( AB\right) +E_2\left( AC\right) +E_2\left( AD\right)
\\ 
&  & +E_2\left( BC\right) +E_2\left( BD\right) +E_2\left( CD\right) \\ 
E_{3t} & = & E_3\left( ABC\right) +E_3\left( ABD\right) \\ 
&  & +E_3\left( ACD\right) +E_3\left( BCD\right) \\ 
E_4 & = & E_4\left( ABCD\right)
\end{array}
\right.
\end{equation}
And from Eqs. (\ref{b}), there is 
\begin{equation}
S\left( \rho _{AB}\right) +S\left( \rho _{AC}\right) +S\left( \rho
_{AD}\right) =2\cdot E_{2t}+3\cdot E_{3t}+3\cdot E_4  \label{d}
\end{equation}
It follows from eq. (\ref{c}) and (\ref{d}) that 
\begin{equation}
E_4=\sum_{i\in \left\{ A,B,C,D\right\} }S\left( \rho _i\right) -\left\{
S\left( \rho _{AB}\right) +S\left( \rho _{AC}\right) +S\left( \rho
_{AD}\right) \right\}  \label{e}
\end{equation}
This is the amount of "essential" $4$-party entanglement contained in the
state $\left| \Psi \right\rangle _{ABCD}$, therefore it must be
non-negative, i.e., 
\begin{equation}
\sum_{i\in \left\{ A,B,C,D\right\} }S\left( \rho _i\right) -\left\{ S\left(
\rho _{AB}\right) +S\left( \rho _{AC}\right) +S\left( \rho _{AD}\right)
\right\} \geq 0  \label{e1}
\end{equation}
Eq. (\ref{e1}) is a property of all the GHZ reducible $4$-party pure states.
Similar results for $m$-party GHZ reducible pure states can follow from the
same argument.

Now let us take the state 
\begin{equation}
\left| \psi \right\rangle _{ABCD}=\frac 12\left\{ \left| 0000\right\rangle
+\left| 0110\right\rangle +\left| 1001\right\rangle -\left|
1111\right\rangle \right\}
\end{equation}
as an example. It's obvious that 
\begin{eqnarray}
S\left( \rho _A\right) &=&S\left( \rho _B\right) =S\left( \rho _C\right)
=S\left( \rho _D\right) =1  \nonumber \\
S\left( \rho _{AB}\right) &=&S\left( \rho _{AC}\right) =2 \\
S\left( \rho _{AD}\right) &=&1  \nonumber
\end{eqnarray}
therefore 
\begin{equation}
E_4=4\times 1-\left( 2+2+1\right) =-1
\end{equation}
This contradicts eq. (\ref{e1}). Thus we have shown that not all $4$-party
pure states are GHZ reducible, so the set of $2$-, $3$-, $4$-party GHZ
states is not a MREGS for $4$-party pure states. This completes the proof of
proposition 1.

Proposition 1 shows that the set of $11$ generalized GHZ states in eq. (\ref
{x1}) is not enough for a MREGS, i.e., the number of members in a MREGS for $%
4$-party pure states must be greater than 11.

\section{GHZ REDUCIBLE TRIPARTITE PURE STATES}

It was known that any bipartite pure state is GHZ reducible, and from
proposition 1 we know that not all $4$-party pure states are GHZ reducible.
It is natural to ask whether all tripartite pure states are GHZ reducible.
The answer of this question is not found yet, however we give some
properties of the GHZ reducible tripartite pure states.

Let us first recall the definitions of the relative entropy of entanglement
and Rains' bound of entanglement. Let the systems A and B be in a joint
state $\rho _{AB}$, the relative entropy of entanglement $E_r\left(
A,B\right) $ is defined by \cite{vprk} 
\begin{eqnarray}
E_r\left( A,B\right)& \equiv& E_r\left( \rho _{AB}\right)  \nonumber \\
&\equiv& \min_{\sigma \in \widetilde{D}}tr_{AB}\left\{ \rho _{AB}\left( \log
_2\rho _{AB}-\log _2\sigma \right) \right\}
\end{eqnarray}
where $\widetilde{D}$ is the set of all disentangled states of the two
systems A and B. Let $\widetilde{P}$ be the set of all bipartite states that
have positive partial transposes (PPT), similarly Rains' bound of
entanglement $B_{\Gamma}$ is defined by \cite{rains} 
\begin{eqnarray}
B_{\Gamma}\left( \rho _{AB}\right) \equiv& \min_{\sigma \in \widetilde{P}%
}tr_{AB}\left\{ \rho _{AB}\left( \log _2\rho _{AB}-\log _2\sigma \right)
\right\}
\end{eqnarray}
It is obvious that $B_{\Gamma}\left( \rho _{AB}\right) \le E_r\left( \rho
_{AB}\right)$ since any separable state is a PPT state \cite{Pe}. Now we
give the following proposition about the relative entropy of entanglement
for a special kind of tripartite pure states.

{\sl Proposition 2: }For the $3$-party pure state 
\begin{equation}
\left| \Phi \right\rangle _{ABC}=\left| \psi \right\rangle _{AB}^{\otimes
m}\otimes \left| \varphi \right\rangle _{AC}^{\otimes n}\otimes \left| \phi
\right\rangle _{BC}^{\otimes l}\otimes \left| \Theta \right\rangle
_{ABC}^{\otimes k}
\end{equation}
where the state $\left| \Theta \right\rangle _{ABC}$ is Schmidt
decomposable, (i.e., $\left| \Theta \right\rangle _{ABC}=\sum_i\sqrt{p_i}%
\left| i\right\rangle _A\left| i\right\rangle _B\left| i\right\rangle _C$)
there is 
\begin{eqnarray}
E_r\left( A,B\right)  &=&m\cdot E_r\left( \left| \psi \right\rangle
_{AB}\right)   \nonumber \\
E_r\left( A,C\right)  &=&n\cdot E_r\left( \left| \varphi \right\rangle
_{AC}\right)  \\
E_r\left( B,C\right)  &=&l\cdot E_r\left( \left| \phi \right\rangle
_{BC}\right)   \nonumber
\end{eqnarray}

If the relative entropy of entanglement is additive, proposition 2 is
obviously true. However, the additivity of the relative entropy of
entanglement has not been proved yet (maybe it is not provable at all), so
this proposition should be proved. A proof of this proposition can be found
in Appendix A, here we prove this proposition by proving the following lemma.

{\it Lemma 1: } For a bipartite pure state $\rho$ and a bipartite separable
state $\rho^{\prime}$, there is $E_r \left(\rho \otimes
\rho^{\prime}\right)=E_r \left( \rho \right)$.

{\it Proof.} On one hand, it is obvious that \cite{vp} 
\begin{equation}
E_r \left(\rho \otimes \rho^{\prime}\right)\le E_r \left( \rho \right).
\label{hh1}
\end{equation}
On the other hand, as a property of $B_{\Gamma}$, there is \cite{rains} 
\begin{equation}
B_{\Gamma}\left( \rho \otimes \rho^{\prime} \right) = B_{\Gamma}\left( \rho
\right) = E_r \left( \rho \right)
\end{equation}
i.e., 
\begin{equation}
E_r \left(\rho \otimes \rho^{\prime}\right) \ge B_{\Gamma}\left( \rho
\otimes \rho^{\prime} \right) = E_r \left( \rho\right)  \label{hh2}
\end{equation}
Thus lemma 1 follows from eqs. (\ref{hh1}) and (\ref{hh2}). From lemma 1,
and the additivity of the relative entropy of entanglement for pure states,
proposition 2 can easily be proved.

{\it LPSW's lemma: }If two $3$-party (Alice,Bob,Claire) quantum states $%
\left| \Psi \right\rangle $ and $\left| \Phi \right\rangle $ are LOCCa
equivalent, then each of the relative entropies of entanglement of $\left|
\Psi \right\rangle $ is equal to the corresponding one of $\left| \Phi
\right\rangle $, i.e., 
\begin{eqnarray}
E_r^{\left| \Psi \right\rangle }\left( A,B\right) &=&E_r^{\left| \Phi
\right\rangle }\left( A,B\right)  \nonumber \\
E_r^{\left| \Psi \right\rangle }\left( A,C\right) &=&E_r^{\left| \Phi
\right\rangle }\left( A,C\right) \\
E_r^{\left| \Psi \right\rangle }\left( B,C\right) &=&E_r^{\left| \Phi
\right\rangle }\left( B,C\right)  \nonumber
\end{eqnarray}

This lemma follows from LPSW's inequality that for any LOCC protocol, the
average increase in $E_r\left( B,C\right) $ is no greater than the average
decrease in the entanglement between Alice and the joint Bob-Claire system.
Detailed discussion can be found in ref. \cite{lpsw}. By this lemma, LPSW
had made quantitative statements about tripartite entanglement, they notice
that there are relations between the one-party entropies and relative
entropies. Here we look more carefully into this issue and extract the
relations of the entropies.

{\sl Proposition 3: }If tripartite pure state $\left| \Psi \right\rangle
_{ABC}$ is GHZ reducible, then there must be 
\begin{eqnarray}
S\left( \rho _A\right) +E_r\left( B,C\right)& =&S\left( \rho _B\right)
+E_r\left( A,C\right)  \nonumber \\
&=&S\left( \rho _C\right) +E_r\left( A,B\right)  \label{f}
\end{eqnarray}
and 
\begin{equation}
\left. 
\begin{array}{c}
S\left( \rho _A\right) \geq E_r\left( A,B\right) +E_r\left( A,C\right) \\ 
S\left( \rho _B\right) \geq E_r\left( A,B\right) +E_r\left( B,C\right) \\ 
S\left( \rho _C\right) \geq E_r\left( A,C\right) +E_r\left( B,C\right)
\end{array}
\right.  \label{g}
\end{equation}
where $S\left( \rho _A\right) $ is the von Neumann entropy of the reduced
density matrix of system A, and $E_r\left( A,B\right) $ is the relative
entropy of entanglement of the systems A+B.

{\sl Proof. }Since $\left| \Psi \right\rangle _{ABC}$ is GHZ reducible,
i.e., 
\begin{equation}
\begin{array}{lll}
&  & \left| \Psi \right\rangle _{ABC} \stackrel{LOCCa}{\rightleftharpoons }
\left| \Phi \right\rangle _{ABC}\equiv \left| EPR\right\rangle
_{AB}^{\otimes E_2(AB)} \nonumber \\ 
&  & \otimes\left| EPR\right\rangle _{AC}^{\otimes E_2(AC)} \otimes \left|
EPR\right\rangle _{BC}^{\otimes E_2(BC)}\otimes \left| GHZ\right\rangle
_{ABC}^{\otimes E_3(ABC)}
\end{array}
\end{equation}
From LPSW's lemma and proposition 2, we have 
\begin{equation}
\left. 
\begin{array}{c}
E_2\left( AB\right) =E_r\left( A,B\right) \\ 
E_2\left( AC\right) =E_r\left( A,C\right) \\ 
E_2\left( BC\right) =E_r\left( B,C\right)
\end{array}
\right.  \label{h}
\end{equation}
From eq. (\ref{h}) and the additivity of the von Neumann entropy, it follows
that 
\begin{equation}
\left. 
\begin{array}{c}
S\left( \rho _A\right) =E_r\left( A,B\right) +E_r\left( A,C\right)
+E_3\left( ABC\right) \\ 
S\left( \rho _B\right) =E_r\left( A,B\right) +E_r\left( B,C\right)
+E_3\left( ABC\right) \\ 
S\left( \rho _C\right) =E_r\left( A,C\right) +E_r\left( B,C\right)
+E_3\left( ABC\right)
\end{array}
\right.  \label{i}
\end{equation}
Since $E_3\left( ABC\right) \geq 0$, proposition 3 follows from eq. (\ref{i}%
).

Eqs. (\ref{h}) and (\ref{i}) are also obtained in ref. \cite{lpsw}. If we
suppose that the relative entropy of entanglement is additive, then eqs. (%
\ref{h}), (\ref{i}) and proposition 3 are obvious results, however, here we
have given a proof of these results without the assumption of additivity.

We do not know whether conditions (\ref{f}) and (\ref{g}) are satisfied by
all tripartite pure states, but it can be shown that eq. (\ref{f}) is
satisfied for the following case.

{\sl Proposition 4: }For the tripartite pure state $\left| \Psi
\right\rangle _{ABC}$, there are 3 reduced density matrixes of two parties, $%
\rho _{AB}$, $\rho _{AC}$ and $\rho _{BC}$, if at least two of them are
separable states, then eq. (\ref{f}) is satisfied.

Proof of proposition 4 is left to Appendix B.

\section{REDUCIBILITY OF TRIPARTITE PURE STATES AND ADDITIVITY OF THE
RELATIVE ENTROPY OF ENTANGLEMENT}

Let Alice (Bob) hold systems 1 and 3 (2 and 4), $\rho _{12}$ ($\rho _{34}$)
be the joint state of the systems 1 and 2 (3 and 4), and let the systems 1+2
be uncorrelated with the systems 3+4, i.e., the overall state of the systems
1+2+3+4 can be written as 
\begin{equation}
\rho _{AB}=\rho _{12}\otimes \rho _{34}
\end{equation}
We would like to have the additivity 
\begin{equation}
E_r\left( A,B\right) =E_r\left( \rho _{AB}\right) =E_r\left( \rho
_{12}\right) +E_r\left( \rho _{34}\right)  \label{j}
\end{equation}
as an important property desired from a measure of entanglement \cite{vp,bpt}%
. The additivity has been proved for the case that both $\rho _{12}$ and $%
\rho _{34}$ are pure states \cite{vp}, for more general cases, it remains a
conjecture.

{\sl Proposition 5: }The relative entropy of entanglement is additive if
each of the two uncorrelated states (i.e., the above states $\rho _{12}$ and 
$\rho _{34}$) can be purified into a GHZ reducible tripartite pure state.

Proposition 5 says that, if there are two GHZ reducible tripartite pure
states $\left| \psi \right\rangle _{125}$ and $\left| \varphi \right\rangle
_{346}$ such that 
\begin{eqnarray}
\rho _{12} &=&tr_5\left\{ \left| \psi \right\rangle _{125}\left\langle \psi
\right| \right\}  \nonumber \\
\rho _{34} &=&tr_6\left\{ \left| \varphi \right\rangle _{346}\left\langle
\varphi \right| \right\}
\end{eqnarray}
then eq. (\ref{j}) is satisfied. This proposition follows directly from
proposition 2. And we give the following proposition as a corollary.

{\sl Proposition 6: }If all the tripartite pure states are GHZ reducible,
then the relative entropy of entanglement is generally additive.

In another word, if we can find a counter-example for the additivity of the
relative entropy of entanglement, then we can make the statement that not
all tripartite pure states are GHZ reducible.

\section{Conclusions}

In the above discussions, it is shown that the set of generalized GHZ states
is not a minimal reversible entanglement generating set, a MREGS for $m$%
-party pure states ($m\ge 4$) generally includes states other than the
generalized GHZ states, for $4$-party pure states, there must be at least 12
member states in a MREGS.

For the GHZ reducible tripartite pure states, there are strong relations
among the relative entropies of entanglement. And the additivity of the
relative entropy of entanglement is shown to be a necessary condition for
all the tripartite pure states to be GHZ reducible.

\section{ACKNOWLEDGMENTS}

We thank Prof. C. H. Bennett and V. Vedral for valuable communications, and
we also thank Prof. Wu Qiang, Dr. Hou Guang, Mr. Zhou Jindong, Huang
minxing, Luo Yifan, Ms. Chen Xuemei for helpful discussions.

\section{Appendix A: Another Proof of proposition 2}

Before the proof, we first state another lemma.

{\sl Lemma 2: }For bipartite quantum state 
\begin{eqnarray}
\rho &=&\sum_{n_1n_2}a_{n_1n_2}\left| \phi _{n_1}\psi _{n_1}\right\rangle
\left\langle \phi _{n_2}\psi _{n_2}\right|  \nonumber \\
& \equiv & \sum_{n_1n_2}a_{n_1n_2}\left| \phi _{n_1}\right\rangle
_A\left\langle \phi _{n_2}\right| \otimes \left| \psi _{n_1}\right\rangle
_B\left\langle \psi _{n_2}\right|
\end{eqnarray}
the relative entropy of entanglement is given by 
\begin{equation}
E_r\left( \rho \right) =-\sum_na_{nn}\log _2a_{nn}-S\left( \rho \right)
\end{equation}
where $\left| \phi _n\right\rangle $ ($\left| \psi _n\right\rangle $) is a
set of orthogonal normalized states of system A (B), $S\left( \rho \right)
\equiv tr_{AB}\left( -\rho \log _2\rho \right) $ is the von Neumann entropy.

This lemma is a extension of Vedral and Plenio's theorem (Theorem 3 in ref. 
\cite{vp}), the proof is similar to that in ref. \cite{vp}, details can be
found in ref. \cite{wz}, this lemma can also follow directly from Rains'
theorem 9 in ref. \cite{rains}.

Now we come to the proof of proposition 2. The following pure states are
written in their Schmidt decomposition form, 
\begin{equation}
\left. 
\begin{array}{c}
\left| \psi \right\rangle _{AB}^{\otimes m}=\sum_i\sqrt{p_i^\psi }\left|
i^{A_1}\right\rangle \left| i^{B_1}\right\rangle \\ 
\left| \varphi \right\rangle _{AC}^{\otimes n}=\sum_i\sqrt{p_i^\varphi }%
\left| i^{A_2}\right\rangle \left| i^{C_1}\right\rangle \\ 
\left| \phi \right\rangle _{BC}^{\otimes l}=\sum_i\sqrt{p_i^\phi }\left|
i^{B_2}\right\rangle \left| i^{C_2}\right\rangle \\ 
\left| \Theta \right\rangle _{ABC}^{\otimes k}=\sum_i\sqrt{p_i^\Theta }%
\left| i^{A_3}\right\rangle \left| i^{B_3}\right\rangle \left|
i^{C_3}\right\rangle
\end{array}
\right.
\end{equation}
where $p_i^\alpha $ ($\alpha =\psi $,$\varphi $,$\phi $,$\Theta $) satisfy
the normalization condition $\sum_ip_i^\alpha =1$, the systems $A_k$ ($B_k$,$%
C_k$)($k=1,2,3$) are held by Alice (Bob, Claire). Since for pure states the
relative entropy of entanglement is additive \cite{vp}, we have 
\begin{equation}
E_r\left( \left| \psi \right\rangle _{AB}^{\otimes m}\right) =m\cdot
E_r\left( \left| \psi \right\rangle _{AB}\right) =-\sum_ip_i^\psi \log
_2p_i^\psi
\end{equation}
Set $\left| \Psi _1\right\rangle _{ABC}=\left| \psi \right\rangle
_{AB}^{\otimes m}\otimes \left| \Theta \right\rangle _{ABC}^{\otimes k}$,
then 
\begin{eqnarray}
\left| \Psi _1\right\rangle _{ABC} &=& \sum_i\sqrt{p_i^\psi }\left|
i^{A_1}\right\rangle \left| i^{B_1}\right\rangle  \nonumber \\
&&\otimes \sum_j\sqrt{p_j^\Theta }\left| j^{A_3}\right\rangle \left|
j^{B_3}\right\rangle \left| j^{C_3}\right\rangle \\
&=&\sum_{ij}\sqrt{p_i^\psi p_j^\Theta }\left| ij\right\rangle _A\otimes
\left| ij\right\rangle _B\otimes \left| j\right\rangle _C  \nonumber
\end{eqnarray}
therefore 
\begin{eqnarray}
\rho _{AB}^{\left| \Psi _1\right\rangle } &\equiv &tr_C\left\{ \left| \Psi
_1\right\rangle _{ABC}\left\langle \Psi _1\right| \right\}  \nonumber \\
&=&\sum_{ii^{^{\prime }}j}\sqrt{p_i^\psi p_{i^{^{\prime }}}^\psi }\cdot
p_j^\Theta \left| ij\right\rangle _A\left\langle i^{^{\prime }}j\right|
\otimes \left| ij\right\rangle _B\left\langle i^{^{\prime }}j\right|
\end{eqnarray}
From lemma 2, it follows that 
\begin{eqnarray}
E_r\left( \rho _{AB}^{\left| \Psi _1\right\rangle }\right)
&=&-\sum_{ij}p_i^\psi p_j^\Theta \log _2\left( p_i^\psi p_j^\Theta \right)
-S\left( \rho _{AB}^{\left| \Psi _1\right\rangle }\right)  \nonumber \\
&=&-\sum_ip_i^\psi \log _2p_i^\psi -\sum_jp_j^\Theta \log _2p_j^\Theta 
\nonumber \\
&&+ \sum_jp_j^\Theta \log _2p_j^\Theta  \label{Ap1} \\
&=&-\sum_ip_i^\psi \log _2p_i^\psi  \nonumber \\
&=&m\cdot E_r\left( \left| \psi \right\rangle _{AB}\right)  \nonumber
\end{eqnarray}

We now come to prove that 
\begin{equation}
E_r\left( \rho _{A_1B_1}\right) =E_r\left( \rho _{A_1B_1}\otimes \rho
_{A_2}\otimes \rho _{B_2}\right)  \label{Ap2}
\end{equation}
It is known that \cite{vp} 
\begin{eqnarray}
E_r\left( \rho _{A_1B_1}\otimes \rho _{A_2}\otimes \rho _{B_2}\right) &\leq&
E_r\left( \rho _{A_1B_1}\right) +E_r\left( \rho _{A_2}\otimes \rho
_{B_2}\right)  \nonumber \\
&=&E_r\left( \rho _{A_1B_1}\right)  \label{Ap3}
\end{eqnarray}
On the other hand, Alice and Bob can perform local unitary transformations
and measurements to transform the state $\rho _{A_1B_1}\otimes \rho
_{A_2}\otimes \rho _{B_2}$ into the state $\rho _{A_1B_1}\otimes \left|
0\right\rangle _{A_2}\left\langle 0\right| \otimes \left| 0\right\rangle
_{B_2}\left\langle 0\right| $, as the relative entropy of entanglement does
not increase under LOCC, there is 
\begin{eqnarray}
E_r\left( \rho _{A_1B_1}\otimes \rho _{A_2}\otimes \rho _{B_2}\right) &\geq
&E_r\left( \rho _{A_1B_1}\otimes \left| 0\right\rangle _{A_2}\left\langle
0\right| \right.  \nonumber \\
&& \left. \otimes \left| 0\right\rangle _{B_2}\left\langle 0\right| \right)
\label{Ap4} \\
&=&E_r\left( \rho _{A_1B_1}\right)  \nonumber
\end{eqnarray}
The last equality is true since there is no limit on the dimension of the
Hilbert space for the systems held by Alice and Bob. Therefore eq. (\ref{Ap2}%
) follows from eq. (\ref{Ap3}) and (\ref{Ap4}).

From eq. (\ref{Ap1}) and (\ref{Ap2}), we have 
\begin{equation}
E_r\left( A,B\right) =E_r\left( \rho _{AB}^{\left| \Phi \right\rangle
}\right) =E_r\left( \rho _{AB}^{\left| \Psi _1\right\rangle }\right) =m\cdot
E_r\left( \left| \psi \right\rangle _{AB}\right)
\end{equation}
The other two equalities in proposition 2 follow from the symmetry of the
state $\left| \Phi \right\rangle _{ABC}$. Thus the proof of proposition 2 is
completed.

\section{Appendix B: Proof of proposition 4}

Let $\rho _{AB}$ and $\rho _{BC}$ be separable states, then 
\begin{equation}
\rho _{BC}=\sum_{i=1}^Mp_i\left| \psi _i^B\right\rangle \left\langle \psi
_i^B\right| \otimes \left| \phi _i^C\right\rangle \left\langle \phi
_i^C\right|  \label{Ap5}
\end{equation}
where $\varepsilon =\left\{ p_i\text{, }\left| \psi _i^B\phi
_i^C\right\rangle \vert i=1,2,\cdots ,M\right\} $ is an ensemble of $\rho
_{BC}$ with the fewest members.

Let us first show that, {\it the states }$\left| \phi _i^C\right\rangle $%
{\it \ in eq.} (\ref{Ap5}) {\it \ can always be chosen to be orthogonal.}

Alice appends an ancilla and performs a local unitary transformation on $%
\left| \Psi \right\rangle _{ABC}$, resulting in 
\begin{equation}
\left| \widetilde{\Psi }\right\rangle _{ABC}=\sum_{i=1}^M\sqrt{p_i}\left|
i^A\psi _i^B\phi _i^C\right\rangle  \label{Ap6}
\end{equation}
where $\left| i^A\right\rangle $ is a set of orthogonal normalized states of
Alice's enlarged system. The Hughston-Joza-Wootters result \cite{hjw}
ensures that this is always possible. The reduced density matrix 
\begin{eqnarray}
\widetilde{\rho _{AB}} &=&tr_C\left( \left| \widetilde{\Psi }\right\rangle
_{ABC}\left\langle \widetilde{\Psi }\right| \right)  \nonumber \\
&=&\sum_{i,j=1}^M\sqrt{p_ip_j}\cdot \left\langle \phi _j^C\right. \left|
\phi _i^C\right\rangle \cdot \left| i^A\right\rangle \left\langle j^A\right|
\otimes \left| \psi _i^B\right\rangle \left\langle \psi _j^B\right|
\label{Ap7}
\end{eqnarray}
is also a separable state, since local unitary transformation by Alice does
not change the entanglement of the two systems A and B, i.e., $\widetilde{%
\rho _{AB}}$ can be written as 
\begin{equation}
\widetilde{\rho _{AB}}=\sum_kp_k\cdot \rho _k^A\otimes \rho _k^B  \label{Ap8}
\end{equation}
Let $P_A$ be any projection acting on the Hilbert space of system A. It is
obvious that the state 
\begin{equation}
\rho _P=\left( P_A\otimes I_B\right) \widetilde{\rho _{AB}}\left( P_A\otimes
I_B\right)  \label{Ap9}
\end{equation}
is also a separable state (except for a normalization factor). Set $%
P_A=\left| m^A\right\rangle \left\langle m^A\right| +\left| n^A\right\rangle
\left\langle n^A\right| $, therefore 
\begin{eqnarray}
\rho _P &=&p_m\left| m^A\right\rangle \left\langle m^A\right| \otimes \left|
\psi _m^B\right\rangle \left\langle \psi _m^B\right|  \nonumber \\
&&+p_n\left| n^A\right\rangle \left\langle n^A\right| \otimes \left| \psi
_n^B\right\rangle \left\langle \psi _n^B\right|  \nonumber \\
&&+\sqrt{p_mp_n}\left\langle \phi _m^C\right. \left| \phi _n^C\right\rangle
\cdot \left| n^A\right\rangle \left\langle m^A\right| \otimes \left| \psi
_n^B\right\rangle \left\langle \psi _m^B\right|  \nonumber \\
&&+\sqrt{p_mp_n}\left\langle \phi _n^C\right. \left| \phi _m^C\right\rangle
\cdot \left| m^A\right\rangle \left\langle n^A\right| \otimes \left| \psi
_m^B\right\rangle \left\langle \psi _n^B\right|  \label{Ap10}
\end{eqnarray}
Let $\left| \psi _n\right\rangle =\alpha \left| \psi _m\right\rangle +\beta
\left| \psi _m^{\bot }\right\rangle $, where $\beta \left| \psi _m^{\bot
}\right\rangle \equiv \left( \left| \psi _n\right\rangle -\left\langle \psi
_m\right. \left| \psi _n\right\rangle \cdot \left| \psi _m\right\rangle
\right) $ is orthogonal to $\left| \psi _m\right\rangle $. Let the states $%
\left| \psi _m\right\rangle $ and $\left| \psi _m^{\bot }\right\rangle $ be
the basis vectors for the Hilbert space of system B, the partial transpose
of $\rho _P$ is written as 
\begin{equation}
\left( \rho _P\right) ^{T_B}=\left( 
\begin{array}{cccc}
p_m & 0 & K\alpha & 0 \\ 
0 & 0 & K\beta & 0 \\ 
K^{*}\alpha ^{*} & K^{*}\beta ^{*} & p_n\cdot |\alpha |^2 & p_n\alpha \beta
^{*} \\ 
0 & 0 & p_n\alpha ^{*}\beta & p_n\cdot |\beta |^2
\end{array}
\right)  \label{Ap11}
\end{equation}
where $K\equiv \sqrt{p_np_m}\left\langle \phi _n^C\right. \left| \phi
_m^C\right\rangle $. The separability of $\rho _P$ ensures the positivity of
its partial transpose $\left( \rho _P\right) ^{T_B}$ \cite{Pe}, this
positivity requires 
\begin{equation}
\left\langle \phi _n^C\right. \left| \phi _m^C\right\rangle =0\text{ or }%
\beta =0  \label{Ap12}
\end{equation}
i.e., for all $i\neq j$, there is 
\begin{equation}
\left| \phi _j^C\right\rangle \bot \left| \phi _i^C\right\rangle \text{ or }%
\left| \psi _j^B\right\rangle =\left| \psi _i^B\right\rangle  \label{Ap13}
\end{equation}

If $\left| \psi _j^B\right\rangle \neq \left| \psi _i^B\right\rangle $, we
have that $\left| \phi _j^C\right\rangle \bot \left| \phi _i^C\right\rangle $%
. And if $\left| \psi _j^B\right\rangle =\left| \psi _i^B\right\rangle \neq
\left| \psi _k^B\right\rangle $, we can always write 
\begin{eqnarray}
&&p_i\left| \phi _i^C\right\rangle \left\langle \phi _i^C\right| +p_j\left|
\phi _j^C\right\rangle \left\langle \phi _j^C\right|  \nonumber \\
&&=p_i^{^{\prime }}\left| \phi _i^{^{\prime }C}\right\rangle \left\langle
\phi _i^{^{\prime }C}\right| +p_j^{^{\prime }}\left| \phi _j^{^{\prime
}C}\right\rangle \left\langle \phi _j^{^{\prime }C}\right|
\end{eqnarray}
where $p_i^{^{\prime }}+p_j^{^{\prime }}=p_i+p_j$ and $\left| \phi
_j^{^{\prime }C}\right\rangle \bot \left| \phi _i^{^{\prime }C}\right\rangle 
$, each of the two states $\left| \phi _j^{^{\prime }C}\right\rangle $ and $%
\left| \phi _i^{^{\prime }C}\right\rangle $ is a linear addition of the two
states $\left| \phi _j^C\right\rangle $ and $\left| \phi _i^C\right\rangle $%
, so, $\left| \phi _j^{^{\prime }C}\right\rangle $ and $\left| \phi
_i^{^{\prime }C}\right\rangle $ are also orthogonal to $\left| \psi
_k^B\right\rangle $. That is to say, we can rewrite $\rho _{BC}$ as 
\begin{equation}
\rho _{BC}=\sum_{i=1}^Mp_i^{^{\prime }}\left| \psi _i^B\right\rangle
\left\langle \psi _i^B\right| \otimes \left| \phi _i^{^{\prime
}C}\right\rangle \left\langle \phi _i^{^{\prime }C}\right|
\end{equation}
where $\left| \phi _i^{^{\prime }C}\right\rangle $ is a set of orthogonal
normalized states of system C. Thus we prove that, {\it the states }$\left|
\phi _i^C\right\rangle ${\it \ in eq.} (\ref{Ap5}) {\it \ can always be
chosen to be orthogonal.}

Then Alice can append an ancilla and perform a local unitary transformation
on $\left| \Psi \right\rangle _{ABC}$ , resulting in 
\begin{equation}
\left| \widetilde{\widetilde{\Psi }}\right\rangle _{ABC}=\sum_{i=1}^m\sqrt{%
p_i^{^{\prime }}}\left| i^A\psi _i^Bi^C\right\rangle
\end{equation}
where $\sum_ip_i^{^{\prime }}=1$ and $\left| i^A\right\rangle $ ($\left|
i^C\right\rangle \equiv \left| \phi _i^{^{\prime }C}\right\rangle $) is a
set of orthogonal normalized states of system A (C), while $\left| \psi
_i^B\right\rangle $ is a set of normalized states of system B, not
necessarily orthogonal. We have that 
\begin{eqnarray}
\rho _{AC}^{^{\prime }} &=&tr_B\left( \left| \widetilde{\widetilde{\Psi }}%
\right\rangle _{ABC}\left\langle \widetilde{\widetilde{\Psi }}\right| \right)
\nonumber \\
&=&\sum_{ij}\sqrt{p_i^{^{\prime }}p_j^{^{\prime }}}\left\langle \psi
_j^B\right| \left. \psi _i^B\right\rangle \cdot \left| i^Ai^C\right\rangle
\left\langle j^Aj^C\right|
\end{eqnarray}

Since local unitary transformation do not change the relative entropy of
entanglement as well as the von Neumann entropies, from lemma 2, we have 
\begin{eqnarray}
E_r\left( A,C\right) &=&E_r\left( \rho _{AC}\right) =E_r\left( \rho
_{AC}^{^{\prime }}\right)  \nonumber \\
&=&-\sum_ip_i^{^{\prime }}\log _2p_i^{^{\prime }}-S\left( \rho _{AC}\right)
\\
&=&H\left\{ p_i^{^{\prime }}\right\} -S\left( \rho _B\right)  \nonumber
\end{eqnarray}
where $H\left\{ p_i^{^{\prime }}\right\} \equiv -\sum_ip_i^{^{\prime }}\log
_2p_i^{^{\prime }}$. Since $\rho _{AB}$, $\rho _{BC}$ are separable states,
there is 
\begin{equation}
E_r\left( A,B\right) =E_r\left( B,C\right) =0
\end{equation}
And 
\begin{eqnarray}
\rho _A^{^{\prime }} &=&tr_C\left( \rho _{AC}^{^{\prime }}\right)
=\sum_ip_i^{^{\prime }}\cdot \left| i^A\right\rangle \left\langle i^A\right|
\nonumber \\
\rho _C^{^{\prime }} &=&tr_A\left( \rho _{AC}^{^{\prime }}\right)
=\sum_ip_i^{^{\prime }}\cdot \left| i^C\right\rangle \left\langle i^C\right|
\nonumber \\
S\left( \rho _A\right) &=&S\left( \rho _A^{^{\prime }}\right)
=-\sum_ip_i^{^{\prime }}\log _2p_i^{^{\prime }}\equiv H\left\{ p_i^{^{\prime
}}\right\} \\
S\left( \rho _A\right) &=&S\left( \rho _C\right) =H\left\{ p_i^{^{\prime
}}\right\}  \nonumber
\end{eqnarray}
Finally we get the result 
\begin{eqnarray}
S\left( \rho _A\right) +E_r\left( B,C\right) =S\left( \rho _B\right)
+E_r\left( A,C\right)  \nonumber \\
=S\left( \rho _C\right) +E_r\left( A,B\right) =H\left\{ p_i^{^{\prime
}}\right\}
\end{eqnarray}

\widetext

\end{document}